\documentclass[runningheads]{llncs}
\usepackage{amsmath,amssymb}

\DeclareMathOperator{\Loss}{\mathcal{L}}
\usepackage{algorithm}
\usepackage{algorithmic}
\usepackage{graphicx}
\usepackage{multirow}
\usepackage{placeins}
\usepackage{hyperref}

\begin{document}

\title{Boundary Distance Loss for Intra-/Extra-meatal Segmentation of Vestibular Schwannoma}

\titlerunning{Boundary Distance Loss for Vestibular Schwannoma Subsegmentation}

\authorrunning{*}
\author{Navodini Wijethilake\inst{1}, 
Aaron Kujawa\inst{1},
Reuben Dorent\inst{1},  Muhammad Asad\inst{1}, Anna Oviedova\inst{2}, Tom Vercauteren\inst{2}, Jonathan Shapey\inst{1,2}}
\authorrunning{N. Wijethilake et al.}
\institute{School of 
BMEIS,
King’s College London, London, United Kingdom
\and
Department of Neurosurgery, King’s College Hospital, London, United Kingdom} 

\maketitle             
\begin{abstract}
Vestibular Schwannoma (VS) typically grows from the inner ear to the brain. It can be separated into two regions, intrameatal and extrameatal respectively corresponding to being inside or outside the inner ear canal. The growth of the extrameatal regions is a key factor that determines the disease management followed by the clinicians. In this work, a VS segmentation approach with subdivision into intra-/extra-meatal parts is presented. We annotated a dataset consisting of 227 T2 MRI instances, acquired longitudinally on 137 patients, excluding post-operative instances. We propose a staged approach, with the first stage performing the whole tumour segmentation and the second stage performing the intra-/extra-meatal segmentation using the T2 MRI along with the mask obtained from the first stage. To improve on the accuracy of the predicted meatal boundary, we introduce a task-specific loss which we call Boundary Distance Loss. The performance is evaluated in contrast to the direct intrameatal extrameatal segmentation task performance, i.e. the Baseline. Our proposed method, with the two-stage approach and the Boundary Distance Loss, achieved a Dice score of $0.8279\pm0.2050$ and $0.7744\pm0.1352$ for extrameatal and intrameatal regions respectively, significantly improving over the Baseline, which gave Dice score of $0.7939\pm0.2325$ and $0.7475\pm0.1346$ for the extrameatal and intrameatal regions respectively.  
%
%
\end{abstract}
\section{Introduction}

Vestibular schwannomas (VS) are benign intracranial tumours that arise from the insulating Schwann cells of the vestibulocochlear nerve. Typically they begin to grow within the internal auditory canal, often expanding the internal auditory meatus (IAM) and extending medially towards the brainstem, causing symptoms ranging from headache, hearing loss and dizziness to speech and swallowing difficulties as well as facial weakness. VS accounts for 8\% of intra-cranial tumours, and is considered the most common nerve sheath tumour in adults \cite{ostrom2021cbtrus}. 

The decisions of tumour management, which can be either active treatment procedures (surgery or radiotherapy) or wait-and-scan strategy, are taken based on the growth patterns of the tumour \cite{prasad2018decision}. Irrespective of the timing or type of treatment, surveillance of the tumour following treatment is required, where consistent and reliable measurements of the tumour are necessary to estimate tumour size and behaviour \cite{connor2021imaging}.
According to the guidelines for reporting results in Acoustic Neuroma, the intrameatal and extrameatal portions of the tumour are required to be distinguished clearly and the largest extrameatal diameter should be used to report the size of the tumour \cite{kanzaki2003new}. Therefore, this specific segmentation of intrameatal and extrameatal regions is an important task in providing a reliable routine for reporting and analysing the growth of VS. 

Routinely, the extraction of largest extrameatal dimension on the axial plane is performed manually by clinicians as there is no automated framework available in current clinical settings. Thus, the measurements extracted, are prone to subjective variability and also, it is a tedious, labour intensive task \cite{shapey2021artificial}.  Therefore, it is essential to develop an AI framework for intra-/extra-meatal segmentation which we can later integrate into clinical settings along with automated size measurement extraction. With this the reproducibility and repeatability can be ensured. Nonetheless, according to previous studies, the volumetric measures are more repeatable than linear measurements extracted from small VS\cite{mackeith2018comparison,varughese2012growth}.  These volumetric measurements can be reliably extracted using the 3D tumour segmentations. 

\paragraph{Related Work.}
Contrast-enhanced T1 weighted MRI (ceT1) and T2 weighted MRI are frequently utilized for VS management. Several AI approaches have been proposed for VS whole tumour segmentation within the past few years. Shapey et al. \cite{shapey2019artificial} have achieved a Dice score of 0.9343 and 0.8825 with ceT1 and T2 modalities respectively, using a 2.5D convolutional neural network (CNN). Dorent et al. \cite{dorent2022crossmoda} proposed the CrossMoDA computational challenge for VS and cochlea segmentation using T2 MRI with domain adaptation. In CrossMoDA the best performing method reached a Dice score of 0.8840 for the VS structure. 

Using this approach, the authors emphasise how T2 weighted imaging may be routinely utilised for surveillance imaging, increasing patient safety by reducing the need to use gadolinium contrast agents. T2 weighted MRIs are also identified as 10 times more cost-effective than ceT1 imaging \cite{connor2021imaging,coelho2018mri}.
In a recent study, Neve et al \cite{neve2022fully} have reported a Dice score of 0.8700 using T2 weighted MRI on an independent test set, where they have also used the whole tumour segmentation to distinguish the intrameatal and extrameatal regions of VS. 

Multi-stage approaches have been proposed to hierarchically segment substructures of brain gliomas~\cite{wang2017automatic}. The authors claim that cascades can reduce overfitting by focusing on specific regions at each stage  while reducing false positives. However, such approaches have not been used for VS-related tasks.

Boundary-based segmentation losses have been developed
to address the issues associated with the overlap-based losses in highly unbalanced segmentation problems \cite{kervadec2019boundary}. 
Hatamizadeh et al \cite{hatamizadeh2019end} have proposed a deep learning architecture, that consists of a separate module that learns the boundary information which aggregates an the edge aware loss, to the semantic loss. In \cite{zhu2019boundary},
the Boundary Weighted Segmentation Loss (BWSL) combines distance map of the ground truth with the predicted probability map in order to make the network more sensitive to the boundary regions. Overall, existing literature on boundary losses seems focused on closed contours and a solution dedicated to specific boundary sections has not yet been adopted. 

\begin{figure}[t!]
\includegraphics[width=\textwidth]{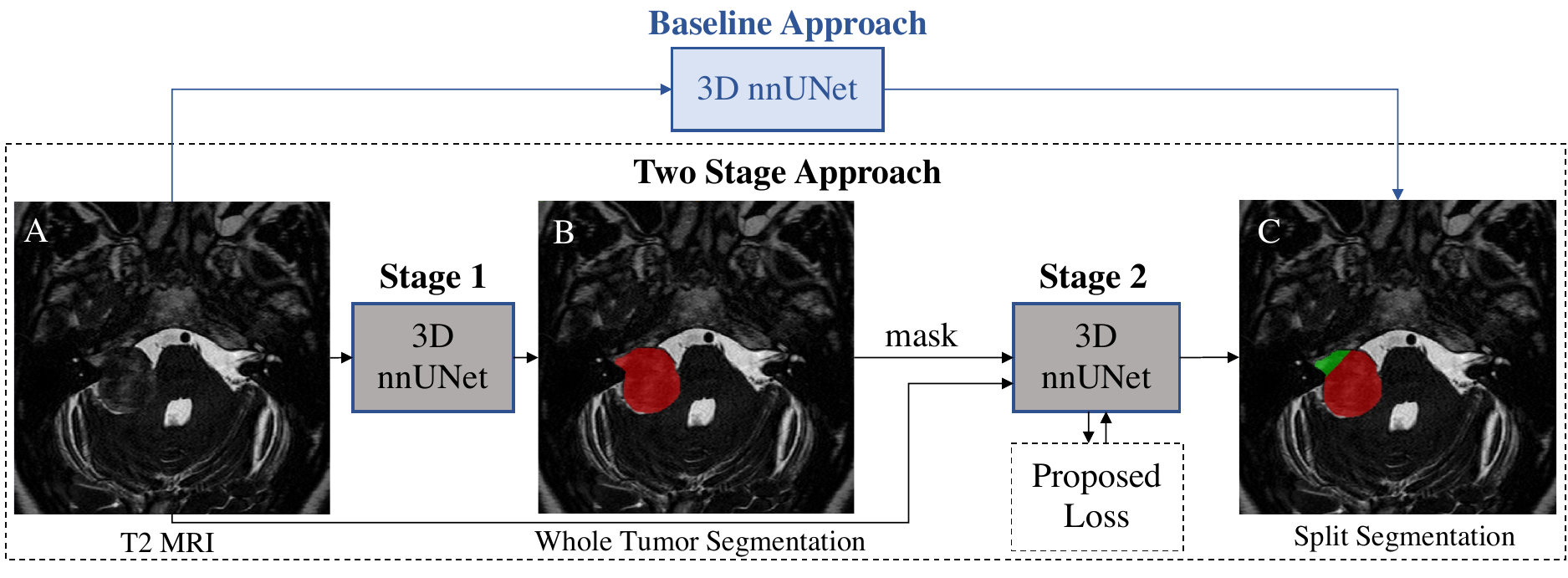}
\caption{The outline of the segmentation task. \textbf{A} shows an axial slice of the T2 MRI volume. \textbf{B} shows the output of stage 1, i.e. the whole tumour segmentation. \textbf{C} shows the output of stage 2, the split segmentation (red label: extrameatal region \& green label: intrameatal region. The two-stage approach is shown within the dotted box and the baseline approach is highlighted in blue above the dotted box. The proposed loss function is used in stage 2. } \label{fig:outline}
\end{figure}

\paragraph{Contributions.}
In this work, we propose a two-stage approach, as illustrated in Figure~\ref{fig:outline}. The first stage performs the whole tumour segmentation and the second stage performs the intra-/extra-meatal segmentation using T2 weighted MRI along with the whole tumour mask obtained from stage 1.
To the best of our knowledge, our study is the first to propose a fully automated learning based approach for intra-/extra-meatal segmentation of VS.
We propose a new Boundary Distance Loss and demonstrate its advantage for learning the boundary between the intra-/extra-meatal regions accurately. We compare the performance of our staged approach and the novel loss function with a baseline, where the split segmentation is performed directly with the T2 weighted MRI volume without the staged approach. Additionally, we also compare the results of stage 2 of the two-stage approach with and without the proposed loss. 

\section{Methods}

\subsubsection{Dataset}
The original dataset consists of MRI scans collected from 165 patients, whose initial MRI scanning was performed during the period February 2006 to January 2017. Further, the follow-up MRI scans were performed until September 2019. The patient cohort was older than 18 years and diagnosed with a single unilateral VS. The whole tumour annotation was performed as an iterative annotation procedure by a specialist MRI-labelling company (Neuromorphometrics, Somerville, Massachusetts, USA), reviewed and validated by a panel of clinical experts that includes a consultant neuroradiologist and a consultant neurosurgeon. 
Subsequently, the intra-/extra-meatal segmentation (split segmentation) was performed for the cases which consisted of segmentations of either contrast-enhanced T1 (ceT1) or T2 weighted MRI modality, by an expert neurosurgeon. Patients who had previously undergone operative surgical treatments were excluded at this stage. For this work, we included only the timepoints with whole tumour and split segmentations on T2 MRI. Thus, our cohort used in this study consists of 227 MRI instances (timepoints) across 137 patients with T2 MRI. The dataset is split into training, validation and testing sets, each with 195, 32 and 56 instances, respectively. We ensure that all timepoints belonging to a single patient are assigned to the same set, i.e. training or validation or testing set. 

This study was approved by the NHS Health Research Authority and Research Ethics Committee (18/LO/0532). Because patients were selected retrospectively and the MR images were completely anonymised before analysis, no informed consent was required for the study. 

\subsubsection{Training: Baseline approach}
For our baseline, we make use of the default 3D full resolution UNet of the nnU-Net framework (3D nnU-Net) \cite{isensee2021nnu} to obtain the intra-/extra-meatal segmentation with the T2 MRI modality as the input for the network. A weighted cross entropy and Dice score losses are used for training. 

\subsubsection{Training: Two-stage nnU-Net approach}
Similarly, we have used the 3D nnU-Net in two-stages sequentially in order to optimize the split segmentation task. All the training was performed on the NVIDIA Tesla V100 GPUs. Each model was trained for 1000 epochs, and the best performing model during validation was used to obtain the inference results. 

\paragraph{Stage 1: Whole tumour segmentation.}
In stage 1, the 3D nnU-Net is used to segment the whole tumour region of VS with the T2 MRI as input. The combined loss of Cross Entropy and Dice score is used in this stage. 

\paragraph{Stage 2: Intra-/Extra-meatal Segmentation.}
In stage 2, the whole tumour mask, in addition to the T2 MRI, is given to the 3D nnU-Net to segment the tumour into intrameatal and extrameatal regions.
For training, the manually annotated masks have been used but during inference, the predicted masks from stage 1 have been used for evaluation. We use a combination of cross entropy, Dice loss and our proposed Boundary Distance Loss detailed below.

\subsubsection{Boundary Distance Loss function}
We define $I \in \mathbb{R}^{H \times W \times D}$ as the T2 MRI volumes with height, width, depth of $H$, $W$, $D$. Any corresponding (probabilistic) binary label map is denoted by $L_{\textrm{label}} \in \mathbb{R}^{H \times W \times D}$. The goal of this proposed loss function is to learn the deviation in the prediction from the actual boundary of intrameatal and extrameatal tumour regions. 

Let's assume that for the three classes (background, intrameatal region and extrameatal region) the prediction map, i.e. the softmax probability maps of the neural network, is denoted by $P = [P_{0}, P_{1}, P_{2}]$. 
The spatial gradients $\nabla_x P_{i}$, $\nabla_y P_{i}$, $\nabla_z P_{i}$ in the $x$, $y$, and $z$ directions provide spatial gradient magnitudes:
\begin{align}
    \lvert \nabla P_{i} \lvert = \sqrt{\nabla_{x}P_{i}^{2} + \nabla_{y}P_{i}^{2} +\nabla_{z}P_{i}^{2}}, \quad i \in \{0,1,2\}
\end{align}

The boundary between the intra-/extra-meatal regions should feature as edges in both corresponding label probability maps. As such, we multiply the magnitudes of the spatial gradients from the corresponding probability maps to achieve an intra-/extra-meatal boundary detector $B_{P}^{1,2}$ from our network predictions:
\begin{equation}
    B_{P}^{1,2} = \lvert \nabla P_{1} \lvert \cdot \lvert \nabla P_{2} \lvert
\end{equation}

Let $L_{B} \in \mathbb{R}^{H \times W \times D}$ denote the ground-truth one-hot encoded binary boundary map between the intra- and extra-meatal tumour regions. The Euclidean distance map $\phi_{B}$ from this ground-truth boundary is defined as,
\begin{equation}
\phi_{B}(u)=\left\{\begin{aligned}
0, \quad & L_{B}(u) = 1 \\
\inf _{v | L_{B}(v)=1}\|u-v\|_{2}, \quad & L_{B}(u) = 0
\end{aligned}\right.
\end{equation}

The PyTorch implementation of the Euclidean distance map is retrieved from the FastGeodis package \footnote{\url{https://github.com/masadcv/FastGeodis}}. 
To promote boundary detections that are close to the true boundary while being robust to changes far away from the true boundary, we compute the average $D_{\phi_{B}}$ negative scaled exponential distance between the detected boundary points and the ground truth:

\begin{equation}
D_{\phi_{B}} = 
\frac{\sum_u B_{P}^{1,2}(u) \exp \left(-\phi_{B}(u)/\tau\right)}{\sum_u B_{P}^{1,2}(u)}
\end{equation}

where $\tau$ is a hyperparameter acting as a temperature term.
We finally compute our Boundary Distance Loss $\Loss_{B}$ by taking the negative logarithm of $D_{\phi_{B}}$

\begin{equation}
    \Loss_{B} =  -\log\left( D_{\phi_{B}} \right)
\end{equation}

To define our complete loss $\Loss$ for training, we combine $\Loss_{B}$ with the cross-entropy ($\Loss_{CE}$) and the Dice loss ($\Loss_{DC}$) weighted by a factor $\gamma$:
\begin{equation}
\Loss = \Loss_{CE} + \Loss_{DC} + \gamma \Loss_{B}
\end{equation}

\subsubsection{Evaluation}

Following the recent consensus recommendations on the selection of metrics for biomedical image analysis tasks \cite{maier2022metrics}, we evaluate our results using the Dice score as the primary overlap-based segmentation metric.
Also following~\cite{maier2022metrics},
the average symmetric surface distance (ASSD) metric is used to measure the the deviation in the prediction, from the actual boundary of intrameatal and extrameatal tumour regions.

\if false
\cite{yeghiazaryan2018family}. 
Let $\hat{L}_{B} \in \mathbb{R}^{H \times W \times D}$ denote the predicted binary map of the boundary between intrameatal and extrameatal tumour regions, as obtained after discretisation of the network outputs $P_i$. The ASSD is defined by:
\begin{align}
\mathrm{~d}(u, L_{\textrm{label}}) &= \min _{v | L_{\textrm{label}}(v)=1} \left\|u-v\right\| \\
\mathrm{ASSD}(\hat{L}_{B},L_{B}) &= \frac{
\sum_{{u | L_{b}(u)=1}} \mathrm{d}(u, \hat{L}_{B})
+
\sum_{{u | \hat{L}_{b}(u)=1}} \mathrm{d}(u, L_{B})
}{\sum_{u}L_{b}(u) +\sum_{u}\hat{L}_{b}(u)}
\end{align}
\fi

\section{Results}
\begin{table}[!ht] \centering
\caption{Inference: Comparison of Dice score. BG: Background, EM: ExtraMeatal, IM: IntraMeatal, WT: Whole tumour, SD: Standard Deviation.}\label{tab:DiceScore}
\begin{tabular}{|l|l|l|l|l|l|l|}
\hline
Method                                              & $\gamma$&       & BG   & EM  & IM  &  WT   \\ \hline
\multirow{2}{*}{nnU-Net (Baseline)}                 &  -    & Mean  &  0.9998  & 0.7939& 0.7475& 0.8813 \\ \cline{2-7}
                                                    &  - &  SD    &  0.0002 & 0.2325 &0.1346&0.0888  \\ \hline
\multicolumn{7}{|l|}{\textit{Two-stage nnU-Net Approach}}  \\ \hline
\multirow{2}{*}{Stage 1: WT Segmentation }          &  -   &  Mean  & 0.9998 &  -  &  -  &   0.9039 \\ \cline{2-7}
                                                    &  -   &   SD   & 0.0002 &  -  &  -  &   0.0470 \\ \hline
\multirow{6}{*}{Stage 2: Split Segmentation} & \multirow{2}{*}{0}    & Mean   & 0.9996 & 0.8068 & 0.7357 & 0.9026 \\ \cline{3-7}
                                             &                       & SD     & 0.0004 & 0.2231 & 0.1522 & 0.0478 \\ \cline{2-7}
                                             & \multirow{2}{*}{0.01} & Mean   & 0.9995 & 0.8072 & 0.7368 & 0.9026 \\ \cline{3-7}
                                             &                       & SD     & 0.0004 & 0.2213 & 0.1524 & 0.0651 \\ \cline{2-7}
                                             & \multirow{2}{*}{0.05} & Mean   & 0.9995 & 0.8087 & 0.7469 & 0.9027 \\ \cline{3-7}
                                             &                       & SD     & 0.0004 & 0.2243 & 0.1431 & 0.0651 \\ \cline{2-7}
                                             & \multirow{2}{*}{0.1}  & Mean   & 0.9996 &  0.8155& 0.7655 & 0.9027 \\ \cline{3-7}
                                             &                       & SD     & 0.0004 &  0.2176& 0.1329 & 0.0476 \\ \cline{2-7}
                                             & \multirow{2}{*}{0.5}  & Mean   & 0.9995 & \textbf{0.8279} & \textbf{0.7744} & 0.9025 \\ \cline{3-7}
                                             &                       & SD     & 0.0004 & 0.2050 & 0.1352 & 0.0478 \\ \hline
\end{tabular}
\end{table}

The quantitative comparison of the Dice score from the inference phase is given in Table \ref{tab:DiceScore}. The Baseline approach gave a Dice score of $0.7939\pm0.2325$ and $0.7475\pm0.1346$ for extrameatal and intrameatal regions. This was improved significantly ($p<0.01$) to $0.8068\pm0.2231$ and $0.7357\pm0.1522$ respectively for extrameatal and intrameatal regions, with the two-stage approach with combined loss of Cross Entropy and Dice Lose. This performance was further enhanced significantly ($p < 0.0001$) with the proposed Boundary Distance Loss ($\Loss_{B}$), which gave a dice score of $0.8279\pm0.2050$ and $0.7744\pm0.1352$ respectively for extrameatal and intrameatal regions with $\gamma = 0.5$.

\begin{table}[t!] \centering
\caption{Inference: Comparison of ASSD metric; $\mathbf{p}_{X}$: $X^{th}$ percentile of the ASSD metric distribution in percentage.}\label{tab:ASSD}
\begin{tabular}{|l|c|c|c|c|}
\hline
Method                                       & $\gamma$ &  Median         &$\mathbf{p}_{75}$&$\mathbf{p}_{25}$ \\\hline
nnU-Net (Baseline)                            &  -       &  0.8384         &   1.1241        & 0.4326 \\ \hline
\multicolumn{5}{|l|}{\textit{Two-stage nnU-Net Approach}}  \\ \hline
Stage 1: Whole tumour Segmentation            &   -      &  -              &   -             &        \\ \hline
\multirow{5}{*}{Stage 2: Split Segmentation} &  0       & 0.8064          &   1.3529        & 0.5690 \\ \cline{2-5}
                                             &  0.01    & 0.8020          &  1.2303         & 0.5351 \\ \cline{2-5}
                                             &  0.05    & 0.7024          & 1.1350          & 0.4830 \\ \cline{2-5}
                                             &  0.1     & 0.7602          &. 0.9597         & 0.4546 \\ \cline{2-5}
                                             &  0.5     &\textbf{0.5417}  &  0.9586         & 0.4181 \\ \hline
\end{tabular}
\end{table}

\begin{figure}[!ht]
\includegraphics[width=1.0\textwidth]{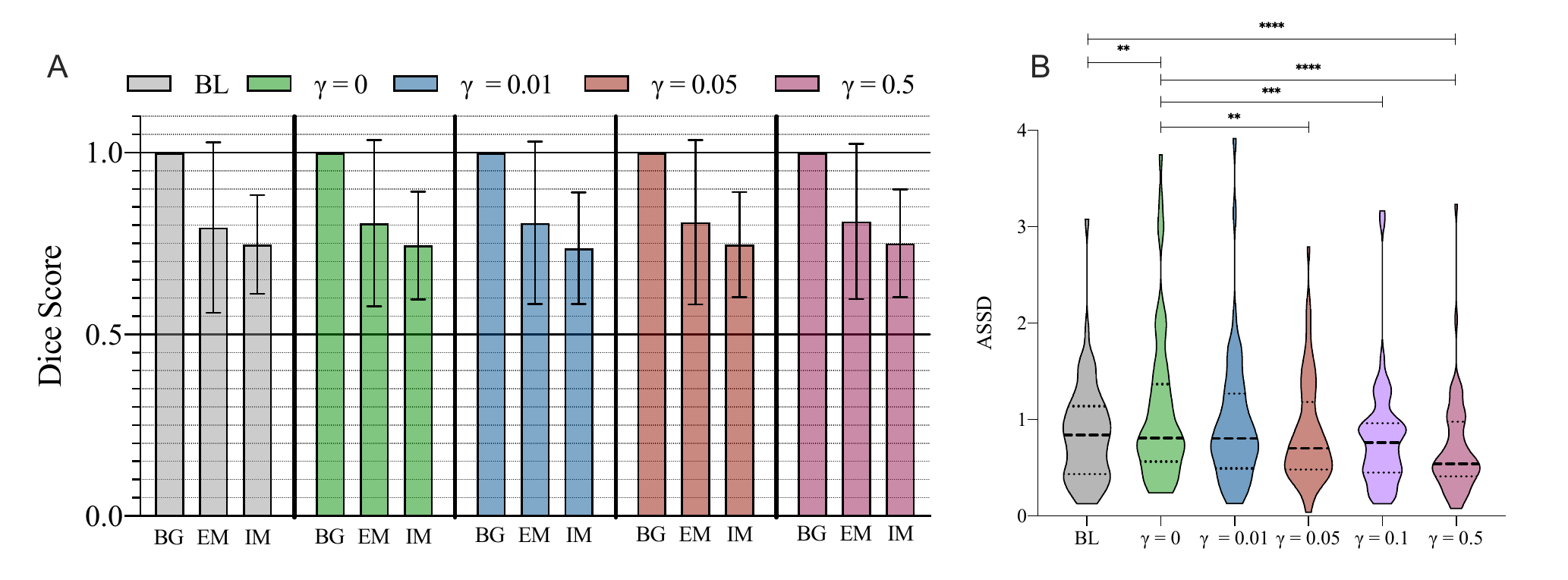}
\caption{\textbf{A} shows the Dice score distribution for the Background (BG), Extrameatal (EM) and Intrameatal (IM) for Baseline method, two-stage approach with $\gamma = 0$, $\gamma = 0.01$, $\gamma = 0.05$ and $\gamma = 0.5$ respectively. \textbf{B} illustrates the ditribution of the ASSD metric for the Baseline method, two-stage approach with $\gamma = 0$, $\gamma = 0.01$, $\gamma = 0.05$ and $\gamma = 0.5$ respectively. } \label{fig:quantitative_results}
\end{figure}

Figure \ref{fig:quantitative_results} shows the distribution of Dice score and the ASSD metric for the Baseline method and two-stage approach for $\gamma = 0$, $\gamma = 0.01$, $\gamma = 0.05$ and $\gamma = 0.5$ respectively. We further performed a two-sided Wilcoxon matched pairs signed-rank test, in which each distribution is considered significantly different from the other distribution when $p < 0.05$.

\begin{figure}[!ht]
\includegraphics[width=0.9\textwidth]{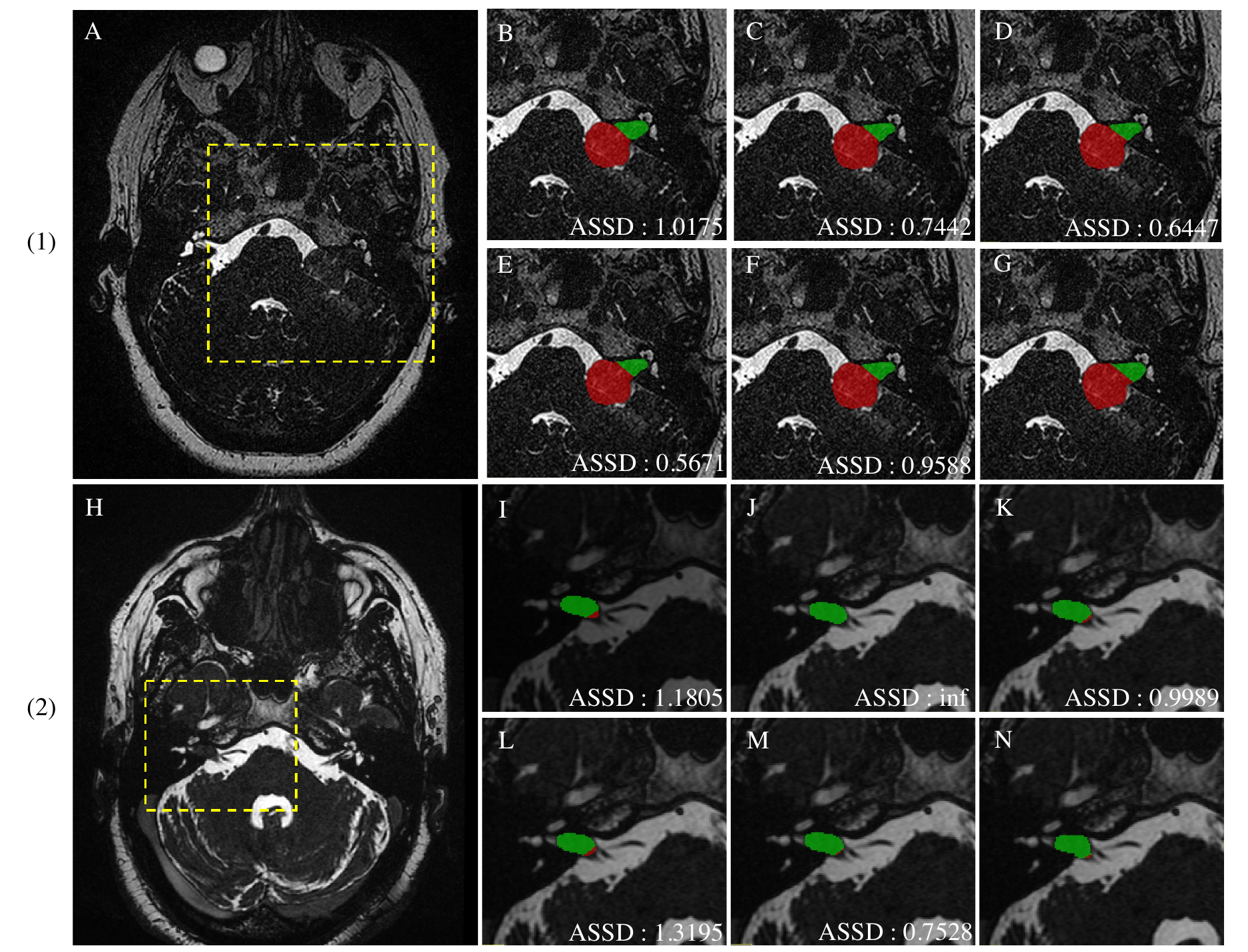}
\caption{\textbf{(1)} \& \textbf{(2)} shows two instances from the testing cohort. The yellow dotted region on \textbf{(1)A}, after segmentation from baseline (direct segmentation) is shown in \textbf{B}. Segmentation output from staged approach with $\gamma = 0$, $\gamma = 0.01$, $\gamma = 0.05$ \& $\gamma = 0.5$ are shown in \textbf{C}, \textbf{D}, \textbf{E}, \textbf{F} respectively. The ground truth is shown in \textbf{G}. Similarly, the yellow dotted region on \textbf{(2)H}, after segmentation from baseline is shown in \textbf{I}. Segmentation output from staged approach with $\gamma = 0$, $\gamma = 0.01$, $\gamma = 0.05$ \& $\gamma = 0.5$ are shown in \textbf{J}, \textbf{K}, \textbf{L}, \textbf{M} respectively. The ground truth is shown in \textbf{N}. The corresponding ASSD metric is shown on each sub figure. } \label{fig:figure_results}
\end{figure}

In Figure \ref{fig:figure_results}, we illustrate qualitative results, i.e. two instances from the testing cohort. The instance (1) gave a high ASSD with baseline and with $\gamma = 0$ for the two-stage approach, compared to the other results obtained with the proposed Boundary Distance Loss. The instance (2) had given a high ASSD metric with the baseline and two-stage approach with the $\gamma = 0$ had given the ASSD value of $\inf$, as it had not distinguished the extrameatal region. The split segmentation obtained with the proposed loss had given comparatively lower ASSD values with $\gamma = 0.01$ and $\gamma = 0.5$ for this instance.

\section{Discussion and conclusion}

In this work, we use a two-stage approach with a novel Boundary Distance Loss for intrameatal and extrameatal segmentation of VS. The two-stage approach with boundary loss shows a promising improvement over the baseline approach of direct intrameatal and extrameatal segmentation.
The two-stage approach helps the model to focus on whole tumour segmentation at the first stage and then, with the mask from stage 1, model can learn the intrameatal and extrameatal separation boundary rigorously during the stage 2. Furthermore, the proposed loss function enhance the split boundary identification by learning the distance between the predicted and the target boundary.  

The results indicate that a low $\gamma$ weight, such as 0.01 or 0.05, on the Boundary Distance Loss does not significantly improve the performance over the zero weight on the proposed loss. However, a significant improvement could be seen with higher $\gamma$ values such as 0.1 or 0.5. In the future, we will fine-tune the $\gamma$ hyper parameter more precisely to assign the most appropriate weight on the proposed Boundary Distance Loss. 

For the training, with the proposed loss the training time for a single epoch increases approximately by two times. In addition, we observed that during inference, if we use the manually annotated mask instead of the mask obtained from the stage 1, the split segmentation improves further. Thus, we can assume if the stage 1 performance can be improved, the stage 2 performance can be enhanced. Another appropriate approach would be training the stage 2 with the predicted masks from the stage 1. 

In conclusion, our method reveals the importance of learning the distance to the boundary in tasks that require distinguishing the boundary precisely. This improvement over the boundary is quite crucial, as it enhances the extraction of features, such as the largest extrameatal diameter from the extrameatal region. Our proposed loss can be used in similar applications that also require a boundary determination from a whole region segmentation.

\section{Acknowledgement}
The authors would like to thank Dr Andrew Worth and Gregory Millington for their contributions to the generation of the segmentation ground truth.
\section{Disclosures}
N. Wijethilake was supported by the UK Medical Research Council [MR/N013700/1] and the King’s College London MRC Doctoral Training Partnership in Biomedical Sciences. This work was supported by Wellcome Trust (203145Z/16/Z, 203148/Z/16/Z, WT106882), EPSRC (NS/A000050/1, NS/A000049/1) and MRC (MC/PC/180520) funding. Additional funding was provided by Medtronic. TV is also supported by a Medtronic/Royal Academy of Engineering Research Chair (RCSRF1819/7/34). SO is co-founder and shareholder of BrainMiner Ltd, UK.

\FloatBarrier
\bibliographystyle{splncs04}
\bibliography{mybib}

\end{document}